\documentclass{article}

\usepackage{arxiv}

\usepackage[utf8]{inputenc} % allow utf-8 input
\usepackage[T1]{fontenc}    % use 8-bit T1 fonts
\usepackage{hyperref}       % hyperlinks
\usepackage{url}            % simple URL typesetting
\usepackage{booktabs}       % professional-quality tables
\usepackage{amsfonts}       % blackboard math symbols
\usepackage{nicefrac}       % compact symbols for 1/2, etc.
\usepackage{microtype}      % microtypography
\usepackage{lipsum}

\usepackage{xfrac}
\usepackage{hhline}
\usepackage{array, multirow}
\usepackage{makecell}
\usepackage{tabu}
\usepackage{amsmath}

\title{Randomized Spectral Sampling for Efficient Simulation of Laser Propagation through Optical Turbulence}

\author{
  Daniel A. Paulson \thanks{Corresponding author.} \\
  Department of Electrical and Computer Engineering\\
  University of Maryland\\
  College Park, MD 20742, USA \\
  \texttt{dpaulson@umd.edu} \\
   \And
  Chensheng Wu \\
  Department of Electrical and Computer Engineering\\
  University of Maryland\\
  College Park, MD 20742, USA \\
   \And
  Christopher C. Davis \\
  Department of Electrical and Computer Engineering\\
  University of Maryland\\
  College Park, MD 20742, USA
}

\begin{document}
\maketitle

\begin{abstract}
We present a new method for the generation of atmospheric turbulence phase screens based on the frequency shift property of the Fourier transform. This method produces low spatial frequency distortions without additional computation time penalties associated with methods using subharmonic subgrids.  It is demonstrated that for simulations of atmospheric turbulence with finite outer scales, the performance of our method with respect to the statistical phase structure function of the screen meets or exceeds other methods with respect to agreement with theory.  We outline small-scale accuracy issues associated with modelling non-Kolmogorov spectral power laws using existing techniques, and propose a solution.  For simulations of long-range propagation through atmospheric optical turbulence, our method provides various advantages over standard methods.
\end{abstract}

\section{Introduction}
\label{sec:introduction}
The split-step propagation method for modeling optical propagation through atmospheric turbulence has been widely used in statistical analysis of beam propagation since its introduction by Fleck, et al. \cite{Fleck1976, Fleck1977}.  This method remains popular in simulations of long-range, linear optical propagation due to computation time advantages associated with using the Fast Fourier Transform (FFT) algorithm to compute the Discrete Fourier Transform (DFT), which are used in both the optical propagation and atmospheric distortion algorithms of the cited method.  In simulations of nonlinear optical propagation phase screens are also widely used \cite{Sprangle2006,Sprangle2011,Palastro2016,Gustafsson2019}, including when supporting studies centering on filamentation \cite{Kandidov_1999,Chin2002,Penano2004,COUAIRON2007,Houard2008,Penano:14}.  However, due to circular shift symmetry and aliasing affects associated with the DFT, significant effort has gone into the development of computational methods which add subharmonic, low-spatial-frequency components to the atmospheric screens \cite{Lane1992,Frehlich2000}.  Additionally, Zernike-polynomial-based methods \cite{Roddier1990} and other creative methods \cite{Burckel13} have been pioneered partly to address this issue, include methods using randomized sampling of the turbulence energy spectrum introduced by Charnotskii \cite{Charnotskii2013}.  However, due to the computational efficiency of leveraging the FFT algorithm, DFT-based methods for phase screen generation remain popular \cite{Xiao2006,Schmidt2010,Nelson2014,Nelson:15,Xiao:16,Voelz2018}.

We present a modified method which exploits the Fourier transform shift theorem \cite{Goodman2017}, which also extends to the DFT \cite{Mitra2001}, in order to include low frequency components in an FFT-centric method in a straightforward manor.  For many applications, this method may provide sufficient phase screen accuracy relative to theory without additional computational penalty associated with subharmonics and other methods.  Additionally the method can be combined with the subharmonic method of Lane, et. al \cite{Lane1992} in order to give very accurate results across a range of spectral models of practical and theoretical importance.  Section \ref{sec:Randomized_FFT-based_Sampling} discusses the basic method and results of use for bounded spectral models where it is well suited.  Section \ref{sec:Hybrid method for use with bounded and unbounded spectral models} discusses the use of the method in concert with subharmonics and other improvements, and includes a detailed analysis of computation time considerations.  Finally, Section \ref{sec:Conclusion} summarizes results and discusses possible applications.  

\section{Randomized FFT-based Sampling} \label{sec:Randomized_FFT-based_Sampling}
\subsection{Algorithm} \label{subsec:Randomized_FFT-based_Sampling_Algorithm}
Central to the study of optical turbulence is the index of refraction structure function, $D_n(\Vec{r})$, defined as \cite{Tartarskii1969}:
\begin{equation}
D_n(\Vec{r}) = \left\langle \left(f\left(\Vec{r}+\Vec{r_o}\right)-f\left(\Vec{r_o}\right)\right)^2 \right\rangle
\label{eq:structure_def}
\end{equation}
As for any real function, the structure function is related to the field's three dimensional energy spectrum, $\Phi_n(\Vec{\kappa})$, by the following relationship \cite{Tartarskii1969}:
\begin{equation}
D_n(\Vec{r}) = 2 \int_{-\infty}^\infty \Phi_n(\Vec{\kappa})\big[1-\cos( \vec{r} \cdot \vec{\kappa})\big] \mathrm{d^3}\vec{\kappa}
\label{eq:structure_spec}
\end{equation}
Traditional phase screen simulations using square grids \cite{Fleck1976,Fleck1977,Lane1992,Frehlich2000,Schmidt2010} approximate the continuous energy spectrum as discrete, and generate complex screens as per:
\begin{equation}
\begin{aligned}
\theta(j,l) = &\sum\limits_{n,m=-M/2}^{M/2-1} {\tilde{c}(n\Delta \kappa_x,m\Delta \kappa_y) exp[2 \pi i(jn+lm)/M]} \\
 = &\sum\limits_{n,m=-M/2}^{M/2-1} {\tilde{c}(n\Delta \kappa_x,m\Delta \kappa_y) exp[i(j n \Delta x \Delta \kappa_x+l m \Delta y \Delta \kappa_y)]}
\label{eq:traditional_phase_screen_formation}
\end{aligned}
\end{equation}
where $i=\sqrt{-1}$, $\Delta x$ and $\Delta y$ are the grid spacings in the $x-$ and $y-$directions, $M$ the number of grid points along each axis, and $\Delta \kappa_x$ and $\Delta \kappa_y$ are the spatial wavenumber grid spacings in the $x-$ and $y-$directions.  We note the relation, $\Delta x \Delta \kappa_x = \Delta y \Delta \kappa_y = 2 \pi / M$.
$\tilde{c}(n\Delta \kappa_x,m\Delta \kappa_y)$ is a random function defined by:
\begin{equation}
\tilde{c}(\beta,\gamma) =  (a+ib) \cdot k \cdot \sqrt{2 \pi \Delta z \Delta \kappa_x \Delta \kappa_y \Phi_n(\beta \hat{e}_x + \gamma \hat{e}_y)}
\label{eq:c_function_def}
\end{equation}
where $\beta$ and $\gamma$ are \textit{dummy}-variables, $k=2 \pi / \lambda$  is the optical wavenumber (with $\lambda$ the wavelength in vacuum), $a$ and $b$ are Gaussian random variables with variances of one, and $\hat{e}_x$ and $\hat{e}_y$ are unit vectors in the $x-$ and $y-$directions.%  It is also important to note that for use with discretized simulations, the use of the function $\Phi_n$ in Eq. \ref{eq:c_function_def} must consider aliasing of spatial angular frequencies greater than $\pi / (M \Delta x)$ to negative frequencies (see supplemental material).

Eq. \ref{eq:traditional_phase_screen_formation} represents a Fourier series using elements which are all harmonic across the spatial domain, creating screens which are periodic \cite{Lane1992}.  Investigating the effect the $\tilde{c}(0,0)$ term has on the summation in Eq. \ref{eq:traditional_phase_screen_formation}, we note that it results in only the addition of a constant phase, \textit{piston} term across all of $\theta(j \Delta x,l  \Delta y)$.  This piston term term does not contribute tip, tilt, focus or defocus effects at any scale, or otherwise contribute to the behavior of the propagating field.  Should $\Phi_n(0,0)$ have a large enough value quantization error \cite{Mitra2001} will result.  $\tilde{c}(0,0)$ is commonly set to zero in practice \cite{Frehlich2000,Schmidt2010}, which avoids these issues. 

In defining a new type of complex phase screen, $\theta_{R}$, we propose a more meaningful use of the point closest to the $\kappa$-space origin by virtue of:
\begin{equation}
\begin{aligned}
\theta_{R}(j,l) = 
\sum\limits_{n,m=-M/2}^{M/2-1} &\tilde{c} \left(n\Delta \kappa_x+\delta \kappa_x,m\Delta \kappa_y+\delta \kappa_y \right) \\
&\hspace{-5.5mm} \cdot exp \left[i \left(j \Delta x  (n \Delta \kappa_x + \delta \kappa_x )+l \Delta y  (m \Delta \kappa_y + \delta \kappa_y) \right) \right]
\label{eq:dither_phase_screen_formation}
\end{aligned}
\end{equation}
where $\delta \kappa_x$ and $\delta \kappa_y$ are random variables described by a uniform distribution bound by $\pm \Delta \kappa_x / 2$ and $\pm \Delta \kappa_y / 2$, respectively.  This offsets the lowest wavenumber grid point away from the origin, along with also translating the rest of the sampling grid in the frequency domain.  By allowing $\tilde{c}(n\Delta \kappa_x+\delta \kappa_x,m\Delta \kappa_y+\delta \kappa_y)$ to define the elements of a matrix, $\tilde{C}(n,m)$, we find that Eq. \ref{eq:dither_phase_screen_formation} is implementable via inverse FFT as per:
\begin{equation}
C(j,l) = M^2 \cdot \mathcal{F}_{2}^{-1}[\tilde{C}(n,m)]
\label{eq:C_def}
\end{equation}
\begin{equation}
\theta_R(j,l) = exp[i(j \Delta x \delta \kappa_x + l \Delta y \delta \kappa_y)] \cdot C(j,l)
\label{eq:dither_phase_screen_FFT}
\end{equation}
where $\mathcal{F}_{2}^{-1}[\ \cdot \ ]$ in Eq. \ref{eq:C_def} denotes the two dimensional inverse FFT operator.  Alternatively, the FFT can be used directly with appropriate conditioning of the $C$ matrix.  

\begin{figure}[ht!]
\centering\includegraphics[width=7.5cm]{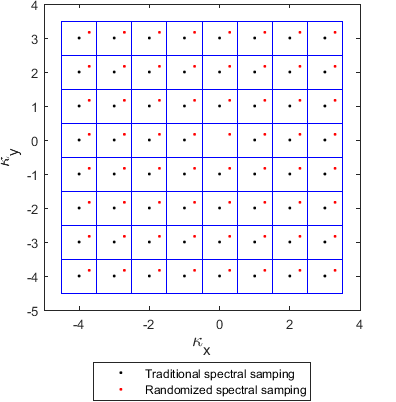}
\caption{Example $\kappa-$space grid partitioning and sampling showing traditional spectral sampling approach versus randomized spectral sampling approach.  Black dots represent traditional sampling points, red dots represent one realization of the randomized sampling approach, and the blue grid lines demarcate the sampling boundaries for the randomized method.}
\label{fig:grid1}
\end{figure}

Fig. \ref{fig:grid1} juxtaposes the sampling methods discussed, where the convention of setting $\tilde{c}(0,0)$ to zero is reflected by the lack of a traditional spectral sampling grid point at the origin.  $\theta_{R}(j,l)$ represents a single complex-number valued phase screen, with the real and imaginary parts therein defining a pair of real-number valued phase screens.  Simulated atmospheric turbulence distortion is applied via multiplication of our complex propagating beam or wave by $exp \left( i \cdot Re\left[ \theta_{R}(j,l) \right] \right)$ or $exp \left( i \cdot Im\left[ \theta_{R}(j,l) \right] \right)$, where $Re$ and $Im$ functions represent taking the real and imaginary parts of an array, respectively.

\begin{figure}[ht!]
\centering\includegraphics[width=8.5cm]{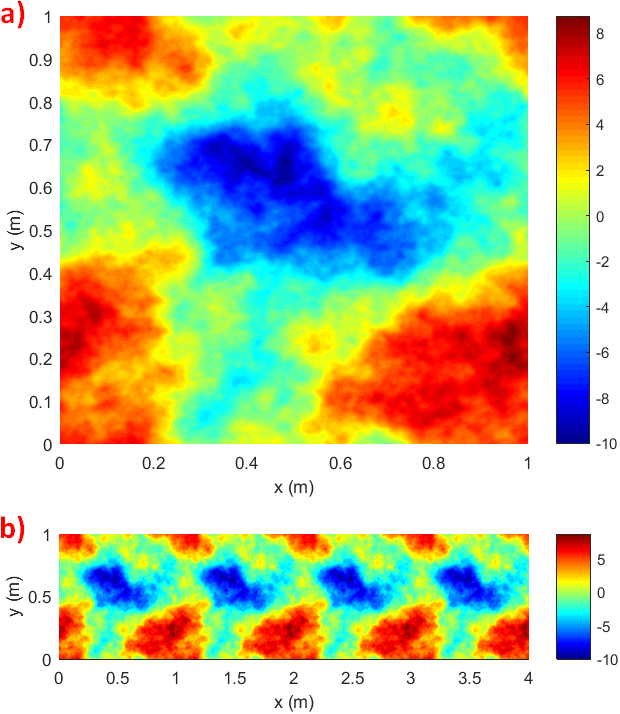}
\caption{a) Phase screen produced using traditional FFT-based algorithm on a 1024 $\times$ 1024 grid; b) Same phase screen as above, repeated four times and placed adjacent to itself in order to illustrate the periodicity associated with traditional FFT-based screens.  The colors shown denote the phase shift of the screen in radians on the simulated propagating wave as per the colorbar.}
\label{fig:undither_phase_screen}
\end{figure}

\begin{figure}[ht!]
\centering\includegraphics[width=8.5cm]{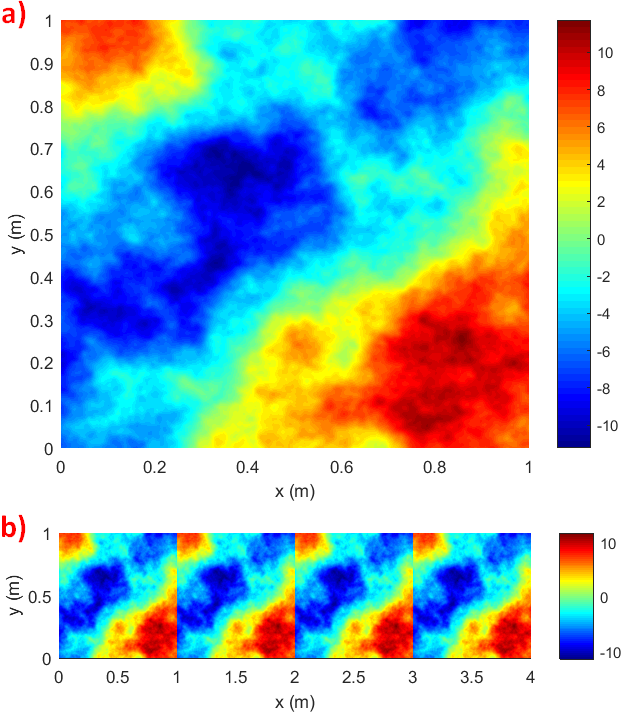}
\caption{a) Phase screen produced using modified FFT-based algorithm on a 1024 $\times$ 1024; b) Same phase screen as above, repeated four times and placed adjacent to itself in order to illustrate the lack of periodicity associated the modified FFT-based algorithm.  As in Fig. \ref{fig:undither_phase_screen}, the color of the screen denotes the phase shift.}
\label{fig:dither_phase_screen}
\end{figure}

$\theta_{R}(j,l)$ and the real valued phase screens it produces no longer exhibit periodicity, and will have domain-wide low spatial frequency distortions.  Additionally, we find the increase in computational delays associated with the use of the algorithm given in Eq. \ref{eq:dither_phase_screen_formation} relative to Eq. \ref{eq:traditional_phase_screen_formation} the context of split-step wave optics simulations to be only 25.4\% – 28.5\% for the grid sizes displayed in this section.  The phase screens shown in Fig. \ref{fig:undither_phase_screen} and Fig. \ref{fig:dither_phase_screen} were generated using the popular approximation to the Hill Spectrum \cite{Hill_1978} developed by Andrews \cite{Andrews_1992,AndrewsBook}, known commonly as the modified atmospheric spectrum.  This spectral model is discussed in detail in subsequent sections.  Note that tip and tilt components can be seen across the $x = 0$ and $y = 0$ axes, respectively, in the screen shown in Fig. \ref{fig:dither_phase_screen}a.  Additionally, Fig. \ref{fig:undither_phase_screen}a displays periodicity \cite{Lane1992}, in that should one circularly shift \cite{Mitra2001} the phase screen in either or both directions no sharp discontinuities would be apparent within the boundaries of the screen.  To help visualize this, we have added Fig. \ref{fig:undither_phase_screen}b and  \ref{fig:dither_phase_screen}b to illustrate the presence and lack of periodicity resulting from the relevant algorithms.

% \section{Use of modified algorithm with bounded spectral models}
\subsection{Results for Bounded Spectral Models} \label{subsec:Randomized_FFT-based_Sampling_Results}
The most widely used three dimension spectral model of atmospheric turbulence is derived from A. Kolmogorov's famous $\sfrac{2}{3}$'s law as \cite{Tartarskii1969,AndrewsBook}: 
\begin{equation}
%\Phi_n(\vec{\kappa}) = \frac{0.033 \cdot C_n^2}{ |\vec{\kappa}| ^{\sfrac{11}{3}}}
%\Phi_n = \frac{0.033 \cdot C_n^2}{ }
\Phi_n(\vec{\kappa}) = {0.033 \cdot C_n^2 \cdot |\vec{\kappa}| ^{-\sfrac{11}{3}}}
\label{eq:Kolmogorov_Spectrum}
\end{equation}
This model is popular due to its simple formulation, and approximate accuracy when the beam statistics of interest are within the inertial subrange of turbulence.  However, this spectrum diverges at the $\kappa$-space origin leading to unphysical properties such as containing infinite energy, divergent covariances, lack of a viscosity driven minimum feature size, and lack of a maximum feature size \cite{AndrewsBook}.  For these reasons, we will refer to this type of spectral model as \textit{unbounded}.  As we will demonstrate in the next section, additional modifications to the FFT-based algorithm may be required to accurately model unbounded spectral models using phase screens, as is the case for Kolmogorov turbulence.

We turn our attention to a practical atmospheric turbulence spectral model which accounts for inner scale, $l_0$, and outer scale, $L_0$, bounds on the inertial subrange, as well as intricacies of the experimentally observed energy spectra at higher spatial frequencies \cite{Champagne1977,williams_paulson_1977,Hill_1978}.  The modified atmospheric spectrum is given by Andrews \cite{Andrews_1992,AndrewsBook}:
%\begin{eqnarray}
%\Phi_n(\kappa) = {0.033 \cdot C_n^2 \cdot \Bigg[ 1 &+& 1.802\bigg(\frac{\kappa}{\kappa_l}\bigg) \\* &-& 0.254\bigg(\frac{\kappa}{\kappa_l}\bigg)^{\sfrac{7}{6}} \Bigg] \cdot \frac{\exp{(-\kappa^2 / \kappa_l^2)}}{(\kappa^2 \kappa_0^2)^{\sfrac{11}{6}}}}
%\label{eq:Modified_Spectrum}
%\end{eqnarray}
%\begin{equation}
%\begin{aligned}
%\Phi_n(\kappa) = 0.033 \cdot C_n^2 \cdot \Bigg[ 1 + 1.802\bigg(\frac{\kappa}{\kappa_l}\bigg) -& %0.254\bigg(\frac{\kappa}{\kappa_l}\bigg)^{\sfrac{7}{6}} \Bigg] \\ \cdot& \frac{\exp{(-\kappa^2 / \kappa_l^2)}}{(\kappa^2 + \kappa_0^2)^{\sfrac{11}{6}}}
%\label{eq:Modified_Spectrum}
%\end{aligned}
%\end{equation}

\begin{equation}
\begin{aligned}
\Phi_n(\vec{\kappa}) = 0.033 \cdot C_n^2 \cdot \textit{f}_n \bigg( \frac{\kappa}{\kappa_l} \bigg) \cdot \frac{\exp{\big(-\kappa^2 / \kappa_l^2\big)}}{\big(\kappa^2 + \kappa_0^2\big)^{\sfrac{11}{6}}}
\label{eq:Modified_Spectrum}
\end{aligned}
\end{equation}
where $\kappa_l=3.3/l_0$, $\kappa_0=2 \pi /L_0$, $\kappa=|\vec{\kappa}|$, and we define the function $f_n$ as:
\begin{equation}
\begin{aligned}
\textit{f}_n (x)  = {1 + 1.802 x - 0.254 x ^ {\sfrac{7}{6}}}
\end{aligned}
\label{eq:f_n_function}
\end{equation}
As this spectral model does not present the same complications as that of Eq. \ref{eq:Kolmogorov_Spectrum}, we refer to this as a \textit{bounded} spectral model.

To assess the accuracy of the revised method, we must designate our metrics of interest.  We had previously defined the refractive index three dimensional structure function, $D_n(\vec{r})$, via the spectral model of interest in Eq. \ref{eq:structure_def}.  The structure function we are interested in, however, is that of an atmospheric phase screen which approximates the cumulative effects of optical propagation through a finite propagation distance, $\Delta z$.  We denote this function as $D_ \theta (\vec{r_\bot})$, where $\vec{r_{\bot}} = x\hat{e_x} + y\hat{e_y}$.  $D_ \theta (\vec{r_\bot})$ is defined by the two dimensional integral over all $\vec{\kappa_\bot}=\kappa_x\hat{e_x} + \kappa_y\hat{e_y}$ as per \cite{Frehlich2000}:
\begin{equation}
D_\theta (\vec{r_\bot}) = 4 \pi k^2 \Delta z \int_{-\infty}^\infty \Phi_n\left(\Vec{\kappa_\bot}\right)\left[1-\cos\left(\vec{r_\bot} \cdot \vec{\kappa_\bot}\right)\right] \mathrm{d^2}\vec{\kappa_\bot}
\label{eq:structure_2D}
\end{equation}
For the modified atmospheric spectrum, though we are aware of closed-form approximations of turbulent structure functions for plane waves applicable to our analysis \cite{Andrews1993}, we have instead developed our theoretical structure function via numerical integration of the equivalent form for isotropic turbulence:
\begin{equation}
D_\theta ( \rho ) = 8 \pi^2 k^2 \Delta z \int_{0}^\infty \kappa_\rho \Phi_n(\kappa_\rho)\big[1-J_0\big(\rho \kappa_\rho)\big] \mathrm{d}\kappa_\rho
\label{eq:structure_2D_Iso}
\end{equation}
where $\rho=|\vec{r_\bot}|$, $\kappa_\rho=|\vec{\kappa_\bot}|$, and $J_0$ denotes the zeroth-order Bessel function of the first kind.

It is well documented that aliasing effects associated with the FFT-based propagation step of the split-step algorithm make parts of the simulation domain unusable \cite{Fleck1976,Schmidt2010,Mansell2007,Coy2005}.  For this reason, a region of interest must be defined, which drives properties of the simulation.  Number of grid points, simulated resolution, as well as the propagation distance between screens must be chosen carefully \cite{Schmidt2010}.  This requires consideration of many factors, including wavelength, coherence lengths, aperture sizes, etc.  As a practical matter, many studies explicitly dedicate half of the $x-$ and $y-$domain of simulation as guard bands to protect against edge effect aliasing \cite{Flatte1993,Ridley1996}.  Additionally, the requirement of grid sizes greater than or equal to twice the size of the limiting apertures (or regions of interest) is explicit in some analysis of simulated propagation using changes of scale between the source and observation planes
\cite{Mansell2007,Coy2005}.  In our own simulations of Gaussian beam propagation \cite{Nelson2014,Wu2018,Paulson2018}, we typically constrain the beam diameter to half the domain of simulation in each $x-$ and $y-$direction in order to avoid edge aliasing effects.  In order to present our results in a simple fashion, we assume that most users would have a region of interest defined by approximately this inner portion of the simulation domain.

Defining the measured $x-$direction structure function along the $M/2^{th}$ row from the $M/4^{th}$ point to the $M/4+j^{th}$ point as $D_x(j\Delta x)$, and $y-$direction structure function along the $M/2^{th}$ column for its corresponding points as $D_y(l\Delta y)$, we can define our percent root mean square (RMS) error metric, $\mathcal{E}$, in terms of the $D_\theta$ defined by Eq. \ref{eq:structure_2D_Iso} via the equations:

\begin{equation}
\mathcal{E}_{x} = \sqrt{\frac{2}{M}\sum_{j=1}^{M/2} \left( \frac{D_x(j\Delta x)-D_\theta(j\Delta x)}{D_\theta(j\Delta x)} \right )^2}
\label{eq:error_x}
\end{equation}
\begin{equation}
\mathcal{E}_{y} = \sqrt{\frac{2}{M}\sum_{l=1}^{M/2} \left( \frac{D_y(l\Delta y)-D_\theta(l\Delta y)}{D_\theta(l\Delta y)} \right )^2}
\label{eq:error_y}
\end{equation}
\begin{equation}
\mathcal{E} = 100\% \times \frac{\mathcal{E}_{x}+\mathcal{E}_{y}}{2}
\label{eq:error_RMS}
\end{equation}
It should also be noted that as part of this study, the diagonal direction structure function was also assessed, with similar results.  However, because the grid diagonals are not orthogonal to the $x-$ and $y-$directions those metrics are not included in our overall statistics.

\begin{figure}[ht!]
\centering\includegraphics[width=9cm]{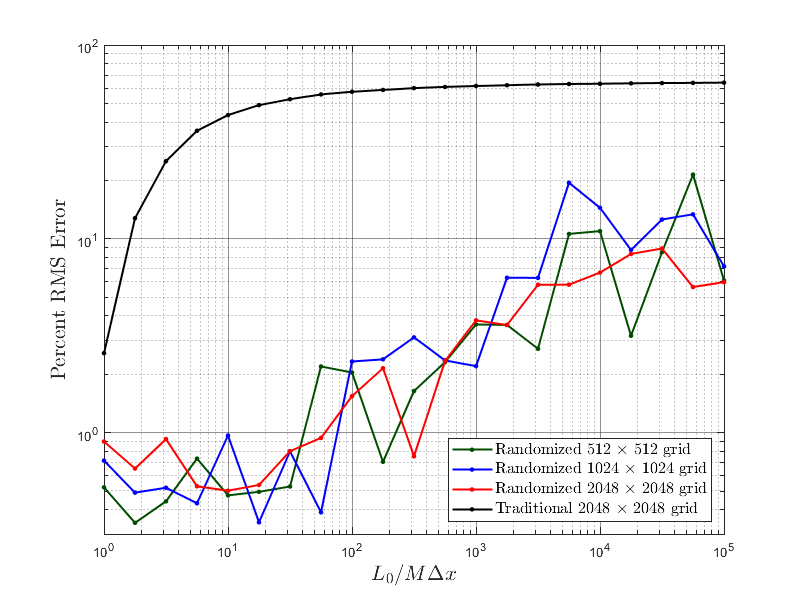}
\caption{RMS error as a percent relative to theory over 50,000 phase screens trials for the simulated domain region of interest as parametrized by the outer scale, $L_0$.  $512 \times 512$, $1024 \times 1024$, $2048 \times 2048$ grid results are shown for randomized method.  For the traditional method the $2048 \times 2048$ grid is shown.}
\label{fig:RMS_Error_Half_Basis}
\end{figure}

\begin{figure}[ht!]
\centering\includegraphics[width=9cm]{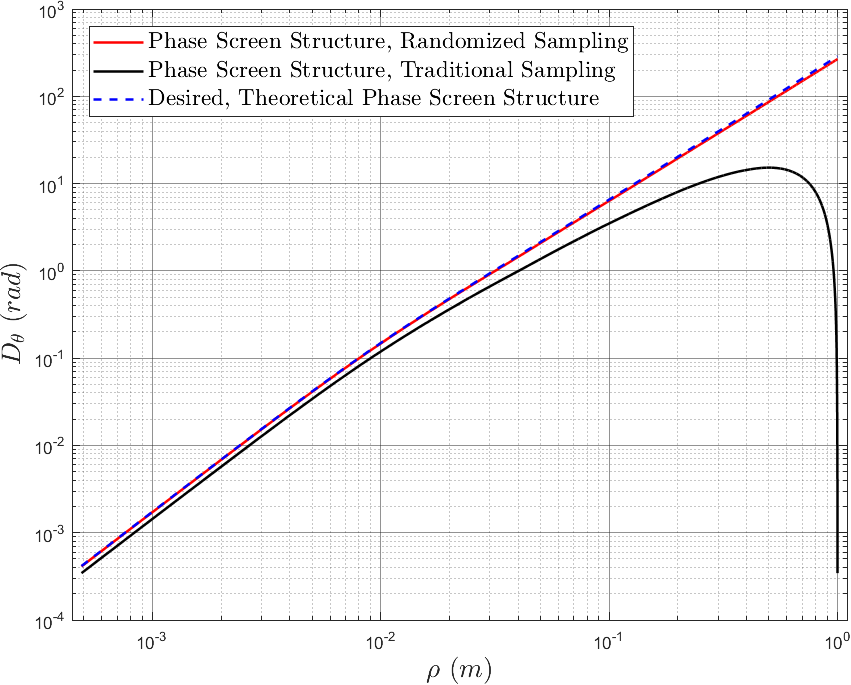}
\caption{Comparison of phase screen structure function versus theory using $2048 \times 2048$ grid, inner scale 1 cm, outer scale 100 km, and effective coherence length of $\rho_o$ = 5 cm.}
\label{fig:structure_example}
\end{figure}

We have found that for the range of outer scale values from one to one thousand times the domain of simulation, the RMS error as a percent assessed over half the simulation domain is constrained to less than 4\%.  More precisely, errors observed over the range $1 \leq L_0/(M \Delta x) \leq 10^{3}$ ranged from 0.34\% - 3.79\%.  For the unrandomized grid, errors ranged from 2.57\% - 61.51\% over the same region.  Fig. \ref{fig:RMS_Error_Half_Basis} displays the Monte-Carlo simulation results over 25,000 complex phase screens.  All data in this study was collected using MATLAB.  Because each complex screen contains a real and imaginary component, and structure function is computed over orthogonal $x-$ and $y-$directions, this simulation set contains 100,000 independent samples per point.  Results are not shown for $512 \times 512$ or $1024 \times 1024$ traditional grids due to overlap of the plotted results, i.e. the results were largely indiscernible from the $2048 \times 2048$ traditional grid results.  Additionally, in order to impress a sense of proportionality upon the reader we have included Fig. \ref{fig:structure_example} which is parametrized by the grid size, outer scale, inner scale, and effective coherence length, $\rho_o$, given for isotropic turbulence and Kolmogorov 11/3rds spectral power laws as \cite{Schmidt2010} $\rho_o = \left(1.46 k^2 \Delta z C_n^2 \right)^{-3/5}$.  On the logarithmic scale, the randomized method follows the theoretical structure function very closely relative to the traditional method.  

\begin{figure}[ht!]
\centering\includegraphics[width=9cm]{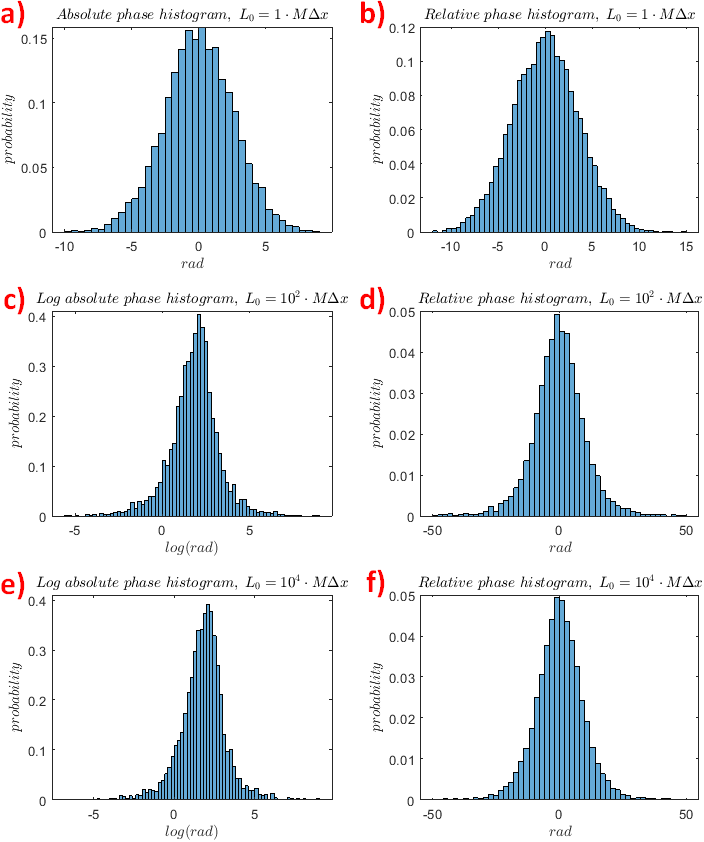}
\caption{Histogram plots of probability density for: a) Absolute phase using outer scale $L_0=1 \cdot M \Delta x$; b) Relative phase over region of interest using outer scale $L_0=1 \cdot M \Delta x$; c) natural logarithm of the magnitude of  absolute phase using outer scale $L_0=10^2 \cdot M \Delta x$; d) Relative phase over region of interest using outer scale $L_0=10^2 \cdot M \Delta x$; e) natural logarithm of the magnitude of  absolute phase using outer scale $L_0=10^4 \cdot M \Delta x$; f) Relative phase over region of interest using outer scale $L_0=10^5 \cdot M \Delta x$.  For each subplot, the probability densities collected over 5,000 phase screens using a $2048 \times 2048$ grid created using the randomized algorithm and modified spectrum.}
\label{fig:histogram}
\end{figure}

Although the structure function is typically the key metric of concern when evaluating phase screen algorithms, the overall probability density function of the resultant screens may be of interest to parties concerned about the overall accuracy of wave optics simulations using the proposed method.  In order to evaluate this, we have run Monte Carlo trials of our proposed method while logging several key parameters:
\begin{enumerate}
    \item Samples of $\theta_{R}(\frac{M}{2},\frac{M}{2})$, the absolute phase of the resultant screens at a central point of the screen.
    \item Samples of $\theta_{R}(\frac{3 M}{4},\frac{M}{2})$ - $\theta_{R}(\frac{M}{4},\frac{M}{2})$, the phase difference across the region of interest in the $x-$direction.
    \item Samples of $\theta_{R}(\frac{M}{2},\frac{3 M}{4})$ - $\theta_{R}(\frac{M}{2},\frac{M}{4})$, the phase difference across the region of interest in the $y-$direction.
\end{enumerate}
Histogram analysis of the results show some interesting characteristics of the collected samples.  For small $L_0$'s, the probability density of the absolute phase was clearly Gaussian in nature, however as $L_0$ was increased this property quickly faded.  Analysis of the natural logarithm of $|\theta_{R}(\frac{M}{2},\frac{M}{2})|$ showed a clear log-normal characteristic as $L_0$ was increased beyond $M \Delta x = 10$ m.  Although a log-normal characteristic of absolute phase may seem undesirable, we do not believe this to be a significant issue with the randomized method as the phase difference histograms maintain approximately zero mean Gaussian distributed characteristics.  Results from the trials for several $L_0$'s are shown in Fig. \ref{fig:histogram}.  We note that the relative phase statistics in the $x-$ and $y-$directions were combined to create the plots on the right hand side of Fig. \ref{fig:histogram}.  Repeating this exercise over several different grid sizes, as well as different separations over which the phase differences were measured, yielded qualitatively similar results.  Since the statistical fluctuations of interest (e.g. scintillation, beam wander, etc.) the of light undergoing atmospheric turbulence distortion are induced by relative phase, we do not believe that the log-normal characteristic of the absolute phase measured will harm the overall statistical properties of the simulated light provided the relative phase maintains zero mean Gaussian characteristics.

In order to demonstrate that the refinement in structure function accuracy using the randomized method reliably improves statistics of propagating light, we have performed wave optics simulations quantifying the angle of arrival (AoA) fluctuations of plane waves propagating through optical turbulence using both our randomized method and the traditional FFT-based method.  Closely following the methodology used by Voelz \cite{Voelz2018} we have simulated 500 nm plane waves propagating through multiple phase screens, using a number of Rytov variances, and collected the aperture averaged AoA's for each trial by focusing the light collected over varying aperture sizes and examining the location of the centroid in the focal plane.  We have deviated from Voelz's method, however, in that instead of fixing the relationship between the propagating plane wave's spatial domain size and aperture diameter, $D$, such that $D / (M \Delta x) = .4$, we have always chosen a spatial domain of  $M \Delta x=1$ m prior to aperturing. Once the propagating light reaches the aperture plane, a focusing transmittance function \cite{VoelzBook} is applied followed by circular aperture functions of several diameters ranging from .5 to 50 cm.  The resulting waves from each aperture function are separately propagated to the focal plane using the angular spectrum propagation method discussed by Schmidt \cite{Schmidt2010} in order to allow for differing spatial domain sizes between the aperture and focal planes.  For all simulation results and theoretical curves shown, the modified spectrum with inner scale $l_0=1$ cm and outer scale $L_0=100$ m was used.

\begin{figure}[ht!]
\centering\includegraphics[width=9cm]{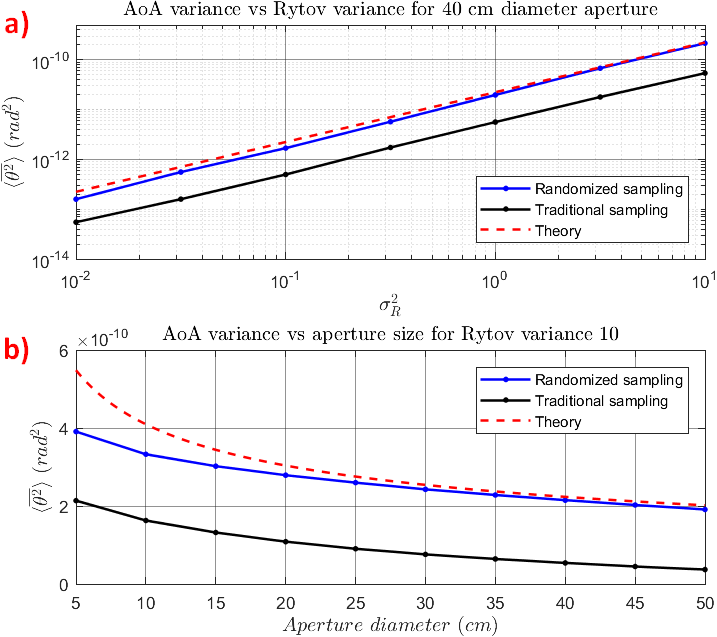}
\caption{a) AoA variance averaged over a 40 cm diameter aperture of a plane wave propagating through turbulence plotted as a function of Rytov variance for wave optics simulations using the randomized method and the traditional method compared to theory; b) Aperture averaged AoA variance of a plane wave propagating through turbulence plotted as a function of aperture diameter for wave optics simulations using the randomized method and the traditional method compared to theory.  All simulations were performed using a $1024 \times 1024$ grid, and 13 equally spaced phase screens over a 2 km propagation distance.}
\label{fig:AoA_var}
\end{figure}

A theoretical expression for the variance of AoA for apertured plane waves is given by Cheon \cite{Cheon:07} as:
\begin{equation}
\langle \overline{\theta^2} \rangle = \pi^2 L \int_{0}^\infty \kappa^3  \Phi_n(\kappa)\left[1 + \frac{2 \pi}{(\kappa f)^2} sin\left( \frac{(\kappa f)^2}{2 \pi} \right) \right] A\left(\frac{D \kappa}{2}\right) \mathrm{d}\kappa
\label{eq:AoA_var}
\end{equation}
where $f=\sqrt{\lambda L}$ is the Fresnel length, and $A(x) = \left( \frac{2J_1(x)}{x} \right)^2$ is the Airy function with $J_1$ denoting the first-order Bessel function of the first kind.  We note that we believe there may have been a typo in \cite{Cheon:07}, and have added a factor of $\pi$ to Eq. 5 in the reference in order to make the equation compatible with later derivations therein.  This theoretical metric has been compared to the results of the wave optics simulations in Fig. \ref{fig:AoA_var}.  In general, we find a marked improvement of results relative to theory when the randomized method is in use for all Rytov variances and aperture sizes.

\section{Hybrid method for use with bounded and unbounded spectral models}
\label{sec:Hybrid method for use with bounded and unbounded spectral models}
\subsection{Core Algorithm}
\label{subsec:Hybrid method for use with bounded and unbounded spectral models, CORE ALGORITHM}
The modified algorithm discussed in the previous section was first investigated with regards to unbounded, anisotropic, non-Kolmogorov spectral models \cite{Paulson2018}.  Therein it was discovered that for structure function power laws greater than the $\sfrac{2}{3}$'s of Kolmogorov the randomized algorithm alone was not sufficient to ensure accurate statistics of observed simulated structure function.  For this reason, we have developed an algorithm utilizing both FFT-based frequency sampling randomization and subharmonic frequency sampling randomization.  We define the following:

\begin{equation}
\begin{aligned}
\theta_{R}(j,l) = 
\sum\limits_{n,m=-M/2}^{M/2-1} &\big(1-\boldsymbol{\delta}[n,m]\big) \cdot \tilde{c}\left(n\Delta \kappa_x+\delta \kappa_x,m\Delta \kappa_y+\delta \kappa_y\right) \\
&\hspace{-7mm} \cdot exp\left[i\left(j \Delta x \left(n \Delta \kappa_x + \delta \kappa_x\right)+l \Delta y \left(m \Delta \kappa_y + \delta \kappa_y\right)\right)\right]
\label{eq:dither_phase_screen_formation_hybrid}
\end{aligned}
\end{equation}

\begin{figure}[ht!]
\centering\includegraphics[width=7.5cm]{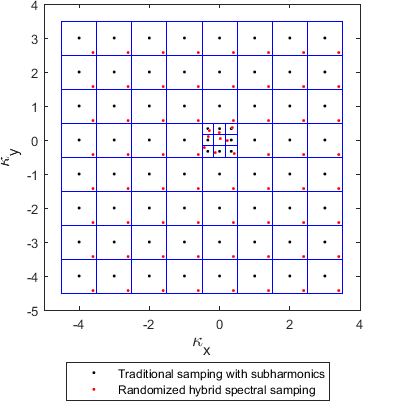}
\caption{Example $\kappa-$space grid partitioning and sampling showing traditional subharmonic sampling approach versus hybrid randomized spectral sampling approach for $N_p=1$. Black dots represent traditional sampling points (including subharmonic expansion), red dots represent one realization of the randomized sampling approach, and the blue grid lines demarcate the sampling boundaries for the randomized method.}
\label{fig:grid2}
\end{figure}

% \begin{equation}
% \begin{aligned}
% \theta_{SH}(j,l) = \sum\limits_{p=1}^{N_p} \sum\limits_{n=-1}^{1} \sum\limits_{m=-1}^{1} &\tilde{c}\bigg(\frac{n\Delta \kappa_x + \delta \kappa_x}{3^{-p}},\frac{m\Delta \kappa_y+\delta \kappa_y}{3^{-p}}\bigg) 
% \\
% & \cdot \big(1-\boldsymbol{\delta}[n,m]\big)
% \\
% & \cdot exp\Bigg[i\Bigg(j \Delta x \bigg(\frac{n\Delta \kappa_x + \delta \kappa_x}{3^{-p}}\bigg) \\ 
% & \ \ \ \ \ \ \ \ \ \ \ \ \ \ \ 
% +l \Delta y \bigg(\frac{m\Delta \kappa_y + \delta \kappa_y}{3^{-p}}\bigg)\Bigg)\Bigg]
% \label{eq:const_phase_screen_formation_hybrid}
% \end{aligned}
% \end{equation}

\begin{equation}
\begin{aligned}
\theta_{out}(j,l,p) = 3^{-2p} \sum\limits_{n,m=-1}^{1} \big(1-&\boldsymbol{\delta}[n,m]\big) \\
&\cdot \tilde{c}\bigg(\frac{n\Delta \kappa_x + \delta \kappa_x}{3^{p}},\frac{m\Delta \kappa_y+\delta \kappa_y}{3^{p}}\bigg) 
\\
& \cdot exp\Bigg[i\bigg(j \Delta x \frac{n\Delta \kappa_x + \delta \kappa_x}{3^{p}} 
\\ 
&\hspace{13mm} +l \Delta y \frac{m\Delta \kappa_y + \delta \kappa_y}{3^{p}}\bigg)\Bigg]
\label{eq:const_phase_screen_formation_hybrid}
\end{aligned}
\end{equation}
\begin{equation}
\begin{aligned}
\theta_{in}(j,l) = 3^{-2(N_p+1)} &\cdot \tilde{c}\bigg(\frac{n\Delta \kappa_x + \delta \kappa_x}{3^{N_p+1}},\frac{m\Delta \kappa_y+\delta \kappa_y}{3^{N_p+1}}\bigg) 
\\
&\cdot exp\Bigg[i\bigg(j \Delta x \frac{n\Delta \kappa_x + \delta \kappa_x}{3^{N_p+1}} \\
&\hspace{13mm}+l \Delta y \frac{m\Delta \kappa_y + \delta \kappa_y}{3^{N_p+1}}\bigg)\Bigg]
\label{eq:inner_phase_screen_formation_hybrid}
\end{aligned}
\end{equation}
In Eq. \ref{eq:inner_phase_screen_formation_hybrid} $N_p$ is the number of subharmonic \textit{constellations} of sampled frequencies (groups of 8 subharmonics chosen from common subgrid boundaries) and $\boldsymbol{\delta}[n,m]$ is the two dimensional discrete Dirac delta function ($\boldsymbol{\delta}[n,m]=1$ for $n=m=0$, otherwise $\boldsymbol{\delta}[n,m]=0$) which we use to ignore the DFT frequency domain origin and the central point of each constellation.  It is very important to note that in Eqs. \ref{eq:const_phase_screen_formation_hybrid} and \ref{eq:inner_phase_screen_formation_hybrid}, we choose a different $\delta \kappa_x$, $\delta \kappa_y$ for each element of the summation (see Fig. \ref{fig:grid2}).  That is, for any index ($n$, $m$) change in in Eq. \ref{eq:const_phase_screen_formation_hybrid} or \ref{eq:inner_phase_screen_formation_hybrid} we choose a new $\delta \kappa_x$, $\delta \kappa_y$ according to a uniform distribution.

The final \textit{hybrid} phase screen, $\theta_{H}$, is given by:

\begin{equation}
\begin{aligned}
\theta_{H}(j,l) = &\theta_R(j,l) + \sum_{p=1}^{N_p} \theta_{out}(j,l,p) + \theta_{in}(j,l)
\label{eq:hybrid_phase_screen_formation}
\end{aligned}
\end{equation}
In Eq. \ref{eq:hybrid_phase_screen_formation}, the summation over $\theta_{out}$ represents the contributions of each of the $N_p$ subharmonic constellations to the phase screen, and $\theta_{in}$ provides the final low-frequency contribution to the phase screen from a spectral sample closest to the origin in $\kappa-$space.  The sampling approach described by Eq. \ref{eq:hybrid_phase_screen_formation} can be visualized by Fig. \ref{fig:grid2} for $N_p=1$.  In Fig. \ref{fig:grid2}, the red dot closest to the origin represents the $\theta_{in}$ spectral sample, and the red dots in the surrounding eight grid partitions represent the $\theta_{out}$ spectral samples associated with the subharmonic constellation.  We have found by choosing the correct number of subharmonic constellations, $N_p$, Eq. \ref{eq:hybrid_phase_screen_formation} yields very accurate results for any reasonable spectral model.  We shall demonstrate results for both bounded and unbounded spectral models later in this section.

\subsection{Addition of White Noise to Phase Screens to Support Subresolution Inner Scales}
\label{subsec:Addition of White Noise to Phase Screens to Support Subresolution Inner Scales}
Since the introduction of generalized spectral models by Kon \cite{Kon1994}, much theoretical work has gone into the study of wave propagation through anisotropic, non-Kolmogorov turbulence defined by unbounded refractive index spectra \cite{Toselli2008,Toselli:11,Andrews2013,Andrews2014,Xiao:16,Wang:16,Beason:18}.  These spectral models are derived from the structure function of refractive index of the form:
\begin{equation}
\begin{aligned}
D_n(x,y,z) = \tilde{C}_n^2 \ \bigg( \frac{x^2}{\mu_x^2} + \frac{y^2}{\mu_y^2} + z^2\bigg)^{\frac{\alpha-3}{2}}
\label{eq:generalized_structure_function}
\end{aligned}
\end{equation}
where $\mu_x$, $\mu_y$ are the anisotropy parameters in the $x-$ and $y-$directions, respectively, $\alpha$ is the three dimensional spectral power law, and $\tilde{C}_n^2$ is the generalized refractive index structure function constant with units $m^{3-\alpha}$.  Although occasionally studies state this structure function model is only valid for $l_0 \ll \sqrt{x^2+y^2+z^2} \ll L_0$ \cite{Andrews2013}, as applied to integrals for calculating second and fourth order beam statistics the inner and outer scales appear as zero and infinity, respectively.  In order for the phase structure function integral definitions of Eqs. \ref{eq:structure_2D} and \ref{eq:structure_2D_Iso} to converge $\alpha$ is typically limited to the range $3 < \alpha < 4$.  Also note that for $\alpha=\sfrac{11}{3}$, Eq. \ref{eq:generalized_structure_function} simplifies to the $\sfrac{2}{3}$'s law of Kolmogorov.  It can be shown \cite{Toselli2008} that Eq. \ref{eq:generalized_structure_function} corresponds to a three dimensional energy spectrum:
\begin{equation}
\begin{aligned}
\Phi_n(\kappa_x,\kappa_y,\kappa_z) = \frac{A(\alpha) \tilde{C}_n^2 \mu_x  \mu_y}{\big(\mu_x^2 \kappa_x^2+\mu_y^2 \kappa_y^2+\kappa_z^2 \big)^{\sfrac{\alpha}{2}}}
\label{eq:generalized_spectrum}
\end{aligned}
\end{equation}
\begin{equation}
\begin{aligned}
A(\alpha) = \frac{\cos{\big(\frac{\pi \alpha}{2}\big)} \Gamma(\alpha-1)}{4 \pi^2}
\label{eq:A_function_def}
\end{aligned}
\end{equation}
where $\Gamma$ denotes the gamma function.

These spectral models do not address practical matters of maximum feature sizes (outer scales) or Kolmogorov microscales (inner scales), where the internal subrange ends and dissipation is the primary form of energy transfer \cite{Ghiaasiaan2011,Fante1975}.  However, these models are useful for studies of non-classical turbulence when the inertial subrange can be approximated as infinite.  For these unbounded cases, great attention has so far been devoted to modeling low spatial frequency components.  In order to explain why this is necessary, we note that insertion of the energy spectrum $\Phi_n(\kappa_x,\kappa_y,0)$ given by Eq. \ref{eq:generalized_spectrum} into the structure function identity given in Eq. \ref{eq:structure_2D} results in an integrand which diverges as $|\vec{\kappa_\bot}|$ approaches zero.   For a more thorough explanation of issues involving use of spectral models which diverge at the zero-frequency point, please see \cite{Paulson2018}.

\begin{figure}[ht!]
\centering\includegraphics[width=9cm]{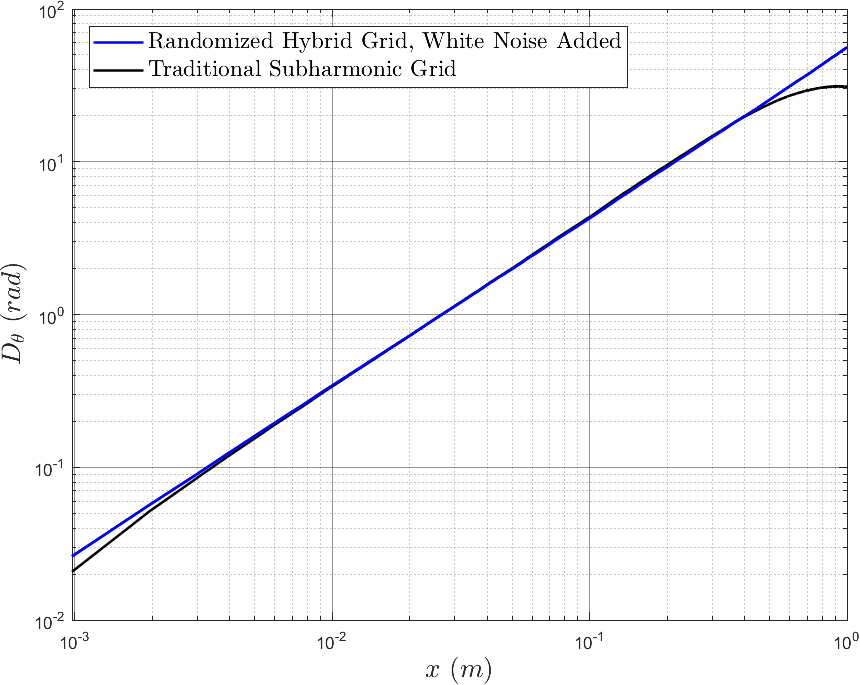}
\caption{$x-$axis structure function of phase screens made with the randomized, hybrid subharmonic algorithm and white noise added, as well as screens using the traditional subharmonic method.  Parameters for the screens are $1024 \times 1024$ grid, $N_p=1$, $\alpha=3.1$, and effective coherence length $\rho_o$ = 5 cm.}
\label{fig:structure_example_white}
\end{figure}

\begin{figure}[ht!]
\centering\includegraphics[width=9cm]{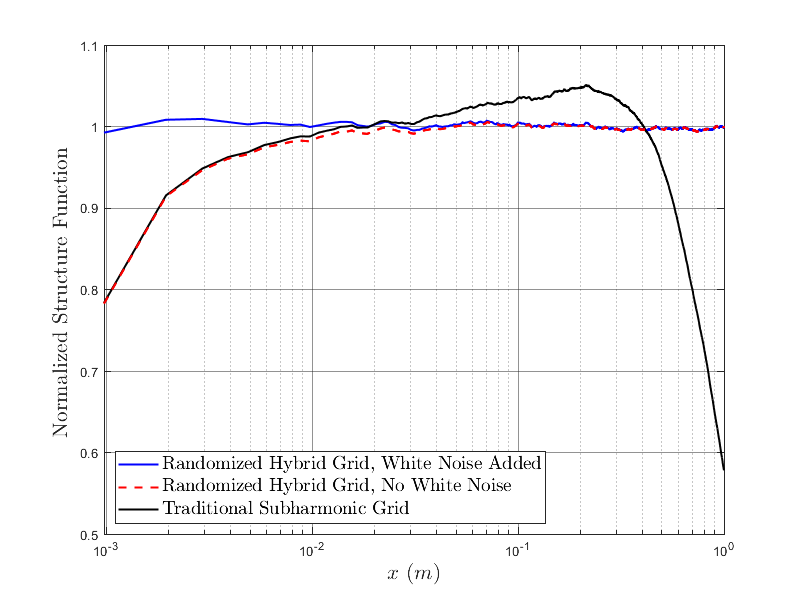}
\caption{$x-$axis structure function normalized by theory of phase screens produced with the randomized, hybrid subharmonic algorithm both with and without white noise added, as well as screens using the traditional subharmonic method.  The parameters used to create this figure are identical to those of Fig. \ref{fig:structure_example_white}}
\label{fig:normalized_structure_example_white}
\end{figure}

Very little emphasis, however, has been placed on high frequency components outside of the simulated $\kappa$-space.  As demonstrated in Figs. \ref{fig:structure_example_white} and \ref{fig:normalized_structure_example_white}, this results in a \textit{sag} of the phase screen structure function relative to theory over small distances.  In order to explain why this sag occurs, we note that the DFT formation of the phase screen algorithm given by Eqs. \ref{eq:traditional_phase_screen_formation}, \ref{eq:dither_phase_screen_formation}, and \ref{eq:dither_phase_screen_formation_hybrid} include spectral energy contributions only within its frequency domain sample space, which we later define explicitly as $K_{in}$.  Spectral energy outside this sample space, $K_{out}$, is not typically included in the discretized simulations, although the structure function integral definition given by Eq. \ref{eq:structure_2D} includes spectral energy across all $\kappa$-space.  We attempt to resolve this problem via addition of white noise to the screen, in order to simulate spectral energy not included in the $\kappa$-space sampling grid or subharmonic subgrids.  Recalling the formation of the structure function in Eq. \ref{eq:structure_2D}, we calculate the variance of the white noise to be added to the screen as per the set of area integrals:
\begin{equation}
\sigma_x^2 = 2 \pi \Delta z k^2 \iint_{K_{out}} \Phi_n(\kappa_x,\kappa_y,0)\big[1-\cos(\Delta x \cdot\kappa_x)\big] \mathrm{d}\kappa_x \mathrm{d}\kappa_y
\label{eq:var_x_def}
\end{equation}
\begin{equation}
\sigma_y^2 = 2 \pi \Delta z k^2 \iint_{K_{out}} \Phi_n(\kappa_x,\kappa_y,0)\big[1-\cos(\Delta y\cdot\kappa_y)\big] \mathrm{d}\kappa_x \mathrm{d}\kappa_y
\label{eq:var_y_def}
\end{equation}
where $K_{out}$ represents the region spanning all of the $\kappa_z=0$ plane, which we define unambiguously via:
\begin{eqnarray}
\label{eq:K}
K = \Big\{ (\kappa_x,\kappa_y): &-&\hspace{-3mm}\infty<\kappa_x< \infty, \\ \nonumber
&-&\hspace{-3mm}\infty<\kappa_y< \infty \Big\} 
\\
\label{eq:Kin}
K_{in} = \Bigg\{ (\kappa_x,\kappa_y): &-&\hspace{-3mm}\Delta \kappa_x \frac{M}{2}<\kappa_x< \Delta \kappa_x \frac{M-1}{2}, \\ \nonumber
&-&\hspace{-3mm}\Delta \kappa_y \frac{M}{2}<\kappa_y< \Delta \kappa_y \frac{M-1}{2} \Bigg\} 
\\
\label{eq:Kout}
K_{out} = \Big\{ (\kappa_x,\kappa_y): &&\hspace{-6mm} (\kappa_x,\kappa_y) \in K \ | \ (\kappa_x,\kappa_y) \notin K_{in} \Big\}
\end{eqnarray}

In practice, the variances of Eqs. \ref{eq:var_x_def} and \ref{eq:var_y_def} can be evaluated numerically as the sum of several integrals.  For the data sets in this article, four integrals per parameter set were used spanning from the each corner of $K_{in}$ to a $|\kappa_x|,|\kappa_y|=\infty$ point in an adjacent quadrant of $\kappa-$space.  Finally, the variances of two white noise processes are calculated as:
\begin{eqnarray}
\label{eq:var_1}
\sigma_1^2 &=& Minimum \left (\sigma_x^2,\ \sigma_y^2 \right ) \\ 
\label{eq:var_2}
\sigma_2^2 &=& \left |\sigma_x^2-\sigma_y^2 \right |
\end{eqnarray}
In order to \textit{whiten} our phase screens, a $M \times M$ matrix of white noise generated variance $\sigma_1^2$ is added to the screen, and is followed by addition of a random number with variance $\sigma_1^2$ across each column (if $\sigma_x^2 > \sigma_y^2$) or row (if $\sigma_y^2 > \sigma_x^2$) of the grid.  For isotropic turbulence $\sigma_x^2=\sigma_y^2$, $\sigma_2^2=0$, and the second step can be negated.  This method ensures the small scale structure function across the $x-$, $y-$, and diagonal directions is improved relative to theory.  All random elements pertaining to white noise are generated using a zero mean Gaussian distribution.

Figures \ref{fig:structure_example_white} and \ref{fig:normalized_structure_example_white} show qualitative results of using this method.  As lower power law $\alpha$ values place a higher portion of their spectral energy at high frequencies, we have chosen to display a power law of $\alpha=3.1$.  In order to reduce the number of independent variables specifying each plot, Figure \ref{fig:structure_example_white} has been given in terms of effective coherence length, $\rho_o$, which for non-Kolmogorov, anisotropic turbulence is given by \cite{Andrews2014}:
\begin{equation}
\rho_o=\left[ \tilde{C}_n^2 k^2 L \frac{\alpha \Gamma(\alpha-1)  \Gamma( - \alpha / 2)}{ 2^\alpha (\alpha - 1) \Gamma(\alpha / 2)} cos \left( \frac{\alpha \pi}{2} \right)  \right]^{\frac{1}{2-\alpha}}
\end{equation}
Additionally, we have included a curve in Fig. \ref{fig:normalized_structure_example_white} showing the phase screen structure function, normalized by theory, produced using the hybrid method but without the addition of white noise. This is intended to demonstrate the positive effects of our white noise algorithm on the relative error at small separations (i.e. near 1\% of the spatial domain and below).

\subsection{Results for Bounded Spectral Models}

\begin{figure}[ht!]
\centering\includegraphics[width=8.5cm]{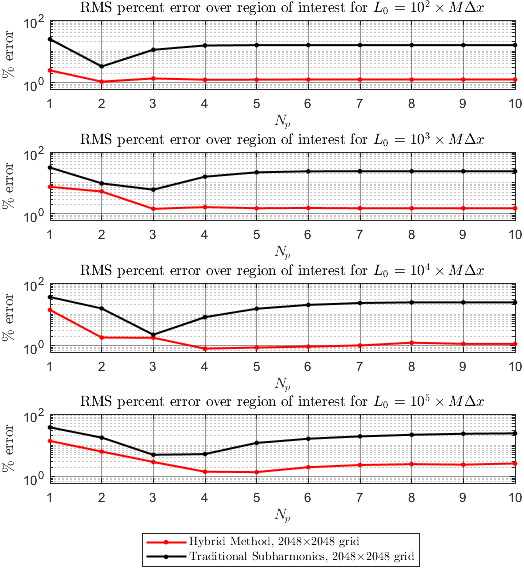}
\caption{RMS error over region of interest computed along $x-$ and $y-$directions for 5,000 phase screens using modified spectrum, with $l_0=M\Delta x / 100$.}
\label{fig:Outer Scale Subh}
\end{figure}

Returning to the modified atmospheric spectrum discussed in Section \ref{sec:Randomized_FFT-based_Sampling}\ref{subsec:Randomized_FFT-based_Sampling_Results}, we observe a marked difference in accuracy of the hybrid method vs the subharmonic method of Frehlich \cite{Frehlich2000}, which we refer to interchangeably as the \textit{traditional subharmonic method}.  We have chosen to compare with this specific subharmonic method, as opposed to other candidates \cite{Lane1992,Johansson1994} due to its improved convergence with theory \cite{Frehlich2000} by virtue of weighting the subharmonic amplitude variances using area integrals of spectral models of interest, as opposed to (non-randomized) spectral samplings.  Results for several values of the outer scale, $L_0$, are shown in Fig. \ref{fig:Outer Scale Subh}, with full results given in Tables \ref{tab:Outer Scale Hybrid Method} and \ref{tab:Outer Scale Trad Method}.  For each case, the size of the outer scale has been set as a factor of the total simulated $x-$, $y-$domain, which was always one meter for this simulation set (i.e. $M \Delta x = 1$ m for all bounded spectrum data sets).  Because it is impractical to present data for all grid sizes, outer scales, and number of subharmonic constellations, we have focused on the $2048 \times 2048$ grid case.  The inner scale for the simulations of this subsection was fixed at $l_0 = M \Delta x / 100$ = 1 cm.  Because the inner scale of interest was many times larger than the resolution of the grid we elected not to include the white noise algorithm in the results of this subsection.

% Defining the measured $x-$direction structure along the $M/2^{th}$ row from the $M/4^{th}$ point to the $M/4+j^{th}$ point as $D_x(j\Delta x)$, and $y-$direction structure along the $M/2^{th}$ column for its corresponding points as $D_y(l\Delta y)$, we can define our error metric, $\mathcal{E}$, in terms of the $D_\theta$ defined by Eq. \ref{eq:structure_2D_Iso} via the equations:

% \begin{equation}
% \mathcal{E}_{x} = \sqrt{\frac{2}{M}\sum_{j=1}^{M/2} \left( \frac{D_x(j\Delta x)-D_\theta(j\Delta x)}{D_\theta(j\Delta x)} \right )^2}
% \label{eq:error_x}
% \end{equation}
% \begin{equation}
% \mathcal{E}_{y} = \sqrt{\frac{2}{M}\sum_{l=1}^{M/2} \left( \frac{D_y(l\Delta y)-D_\theta(l\Delta y)}{D_\theta(l\Delta y)} \right )^2}
% \label{eq:error_y}
% \end{equation}
% \begin{equation}
% \mathcal{E} = 100\% \times \frac{\mathcal{E}_{x}+\mathcal{E}_{y}}{2}
% \label{eq:error_RMS}
% \end{equation}
% It should also be noted that as part of this study, the diagonal direction structure was also assessed, with similar results.  However, because the grid diagonals are not orthogonal to the $x-$ and $y-$directions those metrics are not included in our overall statistics, as they are not independent.

The data in this subsection represents statistics taken from a large sampling of phase screens, along the $\frac{M}{2}+1$ ordered row and column of each screen.  Because the sampling directions are orthogonal the sample set sizes are, essentially, 10,000 trials.  We find that for each case the hybrid method outperforms the traditional subharmonic method, which can be verified by close inspection of Tables \ref{tab:Outer Scale Hybrid Method} and \ref{tab:Outer Scale Trad Method}.  In general, the RMS error over the region of interest in the phase screens can be driven to 3\% or below with the addition of enough subharmonics.  Again comparing Tables \ref{tab:Outer Scale Hybrid Method} and \ref{tab:Outer Scale Trad Method}, we find that the error ratio of the hybrid method with the subharmonic method can be as low as 13.3\%.  

\begin{table}
\begin{tabular}{ |c||p{2.6mm}|p{2.6mm}|p{2.6mm}|p{2.6mm}|p{2.6mm}|p{2.6mm}|p{2.6mm}|p{2.6mm}|p{2.6mm}|p{2.6mm}|p{2.6mm}|}
\hline
- &\multicolumn{10}{c|}{\textbf{Number of Subharmonic Constellations, }$\boldsymbol{N_p}$} \\
\hline
$\boldsymbol{L_0}$ & \textbf{1}  & \textbf{2}  & \textbf{3}  & \textbf{4}  & \textbf{5}  & \textbf{6}  & \textbf{7}  & \textbf{8}  & \textbf{9}  & \textbf{10} \\
\hline
% \hhline{|=||=|=|=|=|=|=|=|=|=|=|=|}
$\boldsymbol{10^{ \ 0}}$ & 1.7 & 1.7 & 1.7 & 1.7 & 1.7 & 1.7 & 1.7 & 1.7 & 1.7 & 1.7 \\
$\boldsymbol{10^{ \ .5}}$ & 2.8 & 3.0 & 3.0 & 3.0 & 3.0 & 3.0 & 3.0 & 3.0 & 3.0 & 3.0 \\
$\boldsymbol{10^{ \ 1}}$ & 2.4 & 2.7 & 2.7 & 2.6 & 2.6 & 2.6 & 2.6 & 2.6 & 2.6 & 2.6 \\
$\boldsymbol{10^{ \ 1.5}}$ & 2.8 & 3.4 & 2.9 & 2.7 & 2.7 & 2.7 & 2.7 & 2.7 & 2.7 & 2.7 \\
$\boldsymbol{10^{ \ 2}}$ & 2.5 & 1.1 & 1.4 & 1.2 & 1.3 & 1.3 & 1.3 & 1.3 & 1.3 & 1.3 \\
$\boldsymbol{10^{ \ 2.5}}$ & 2.6 & 1.1 & 1.3 & 1.1 & 1.3 & 1.3 & 1.3 & 1.3 & 1.3 & 1.3 \\
$\boldsymbol{10^{ \ 3}}$ & 7.5 & 5.3 & 1.4 & 1.6 & 1.5 & 1.5 & 1.5 & 1.5 & 1.5 & 1.5 \\
$\boldsymbol{10^{ \ 3.5}}$ & 10 & 1.6 & 1.5 & 1.5 & 1.1 & 1.3 & 1.4 & 1.3 & 1.3 & 1.3 \\
$\boldsymbol{10^{ \ 4}}$ & 14 & 1.8 & 1.8 & 0.8 & 0.9 & 0.9 & 1.0 & 1.2 & 1.2 & 1.1 \\
$\boldsymbol{10^{ \ 4.5}}$ & 7.4 & 7.0 & 1.8 & 1.6 & 1.9 & 1.9 & 1.5 & 1.5 & 1.6 & 1.6 \\
$\boldsymbol{10^{ \ 5}}$ & 14 & 6.4 & 3.0 & 1.4 & 1.4 & 2.0 & 2.4 & 2.5 & 2.4 & 2.6 \\
\hline
\end{tabular}
\caption{Hybrid method percent RMS error compared to theory over region of interest computed using 5,000 phase screens' $x-$ and $y-$axes for $2048 \times 2048$ grid size, parameterized by outer scale, $L_0$, as well as the number of subharmonic constellations, $N_p$.  The $L_0$'s given in the first column have units of meters, and the simulated spatial domain was a one meter by one meter area.}
\label{tab:Outer Scale Hybrid Method}
\end{table}
 
\begin{table}
\begin{tabular}{ |c||p{2.6mm}|p{2.6mm}|p{2.6mm}|p{2.6mm}|p{2.6mm}|p{2.6mm}|p{2.6mm}|p{2.6mm}|p{2.6mm}|p{2.6mm}|p{2.6mm}|}
\hline
 - &\multicolumn{10}{c|}{\textbf{Number of Subharmonic Constellations, }$\boldsymbol{N_p}$} \\
\hline
$\boldsymbol{L_0}$ & \textbf{1}  & \textbf{2}  & \textbf{3}  & \textbf{4}  & \textbf{5}  & \textbf{6}  & \textbf{7}  & \textbf{8}  & \textbf{9}  & \textbf{10} \\
\hline
% \hhline{|=||=|=|=|=|=|=|=|=|=|=|=|}
$\boldsymbol{10^{\ 0}}$ & 2.1 & 2.2 & 2.2 & 2.2 & 2.2 & 2.2 & 2.2 & 2.2 & 2.2 & 2.2 \\
$\boldsymbol{10^{\ 0.5}}$ & 5.9 & 4.8 & 4.8 & 4.8 & 4.8 & 4.8 & 4.8 & 4.8 & 4.8 & 4.8 \\
$\boldsymbol{10^{\ 1}}$ & 5.4 & 4.5 & 4.9 & 4.8 & 4.8 & 4.8 & 4.8 & 4.8 & 4.8 & 4.8 \\
$\boldsymbol{10^{\ 1.5}}$ & 14  & 7.6 & 13 & 14 & 14 & 14 & 14 & 14 & 14 & 14 \\
$\boldsymbol{10^{\ 2}}$ & 24 & 3.3 & 11 & 15 & 16 & 16 & 16 & 16 & 16 & 16 \\
$\boldsymbol{10^{\ 2.5}}$ & 30 & 7.9 & 8.5 & 17 & 20 & 20 & 20 & 20 & 20 & 20 \\
$\boldsymbol{10^{\ 3}}$ & 31 & 9.7 & 6.1 & 16 & 22 & 24 & 24 & 24 & 24 & 24 \\
$\boldsymbol{10^{\ 3.5}}$ & 35 & 15 & 2.6 & 8.8 & 17 & 22 & 23 & 23 & 23 & 23 \\
$\boldsymbol{10^{\ 4}}$ & 36 & 15 & 2.2 & 8.2 & 15 & 20 & 23 & 24 & 24 & 24 \\
$\boldsymbol{10^{\ 4.5}}$ & 36 & 16 & 2.7 & 7.6 & 14 & 19 & 22 & 24 & 25 & 25 \\
$\boldsymbol{10^{\ 5}}$ & 37 & 18 & 5.0 & 5.3 & 12 & 16 & 19 & 22 & 23 & 24 \\
\hline
\end{tabular}
\caption{Traditional method percent RMS error compared to theory over region of interest in similar configuration as described in Table \ref{tab:Outer Scale Hybrid Method}.}
\label{tab:Outer Scale Trad Method}
\end{table}

\subsection{Results for Unbounded Spectral Models}
\label{subsec:Results for Unbounded Spectral Models}

We also wish to assess the accuracy of our revised algorithm for the cases of generalized anisotropic, non-Kolmogorov turbulence spectra discussed in Section \ref{sec:Hybrid method for use with bounded and unbounded spectral models}\ref{subsec:Addition of White Noise to Phase Screens to Support Subresolution Inner Scales}.  Although in the previous Section the theoretical structure function metric was computed using numeric integration, in this case simple closed-form solutions exist.  By applying a change of variable to Eq. \ref{eq:structure_2D} and we obtain:
\begin{equation}
D_\theta(x,y)=8\pi^2k^2 A(\alpha) B(\alpha) \tilde{C}_n^2 \Delta z \left (\frac{x^2}{\mu_x^2} + \frac{y^2}{\mu_y^2}\right )^\frac{\alpha-2}{2}
\end{equation}
$B(\alpha)$ is defined in terms of a Bessel function integral identity \cite{NIST:DLMF}:
\begin{equation}
\begin{aligned}
B(\alpha) &= \int_0^\infty \left (1-J_0(\kappa_\rho)\right )\kappa_\rho^{(1-\alpha)} \ \mathrm{d}\kappa_\rho \\
&= \frac{\pi \cdot \sec \left (\pi\frac{\alpha-3}{2}\right)}{2^{\alpha-1}\Gamma^2 \left ( \frac{\alpha}{2} \right )}
\end{aligned}
\end{equation}
where \textit{sec} denotes the secant function.  Substituting $D_\theta(j\Delta x,0)$ and $D_\theta(0,l\Delta y)$ for the theoretical error expressions of Eqs. \ref{eq:error_x} and \ref{eq:error_y}, respectively, allows us to again use Eq. \ref{eq:error_RMS} as our error metric.  

\begin{figure}[ht!]
\centering\includegraphics[width=8.5cm]{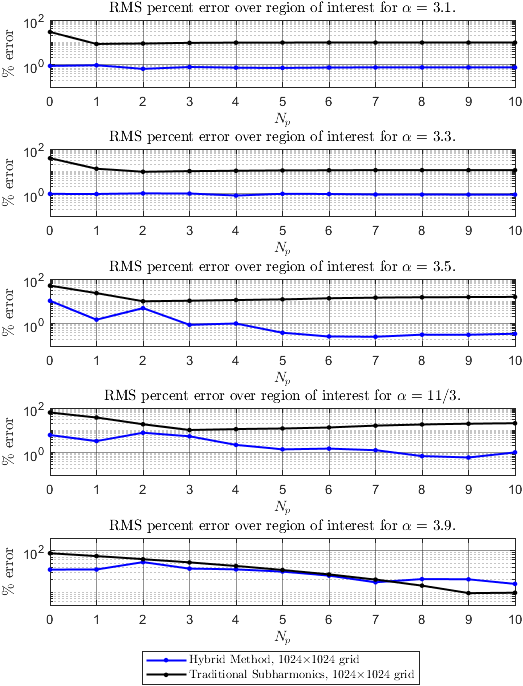}
\caption{RMS error over region of interest computed along $x-$ and $y-$directions of 50,000 phase screens for $1024 \times 1024$ grid using various spectral power laws ($\alpha's$) and number of subharmonic constellations, $N_p$.  $\mu_x=1$ and $\mu_y=2$ for all data points.}
\label{fig:Aniso Subh 1024}
\end{figure}

Due to the combination of our assessment of error as a ratio relative to a theory, as well as the scale invariance \cite{Lane1992} of this section's turbulence models, the specific $\tilde{C}_n^2$ and domain lengths, $M\Delta x$ and $M\Delta y$, do not affect results.  The results are, however, sensitive to the $\alpha$ in use and number of subharmonic constellations.  Figure \ref{fig:Aniso Subh 1024} displays a comparison of the hybrid method, including the addition of white noise, versus the Frehlich subharmonic method for several $\alpha$'s over 50,000 independent, $1024 \times 1024$ resolution phase screens using numbers of constellations between zero and ten.  We note that data was also collected for $\alpha$'s of 3.2, 3.4, 3.6, 3.7, and 3.8 up to $N_p=10$ however that data is not plotted.  Table \ref{tab:min_error_1024} summarizes the minimum errors observed in testing for both schemes, as well as associated $N_p$.  We have observed that our hybrid method outperforms the Frehlich subharmonic method on the $1024 \times 1024$ grid for any number of subharmonic constellations, except for $\alpha=3.9$.  Noting that our revised method falls short for the $\alpha=3.9$ case (at least for $N_p \leq 10$), comparing the minimum $\mathcal{E}$ observed using each method for $\alpha=3.1$ to $3.8$ we note that the average ratio of our hybrid method's $\mathcal{E}$ to that of the Frehlich method is 7.4\%.

% Full results for $\alphas$'s of 3.1 to 3.9 in steps of .1, as well as the Kolmogorov case $\alpha=\sfrac{11}{3}$ are given as Supplemental Material in \ref{sec:Supplemental Material}.
 
\begin{figure}[ht!]
\centering\includegraphics[width=8.5cm]{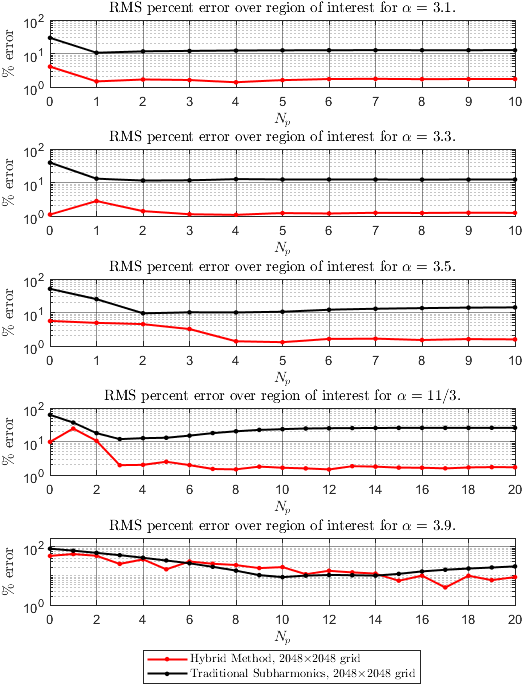}
\caption{RMS error over region of interest computed along $x-$ and $y-$directions of 5,000 phase screens for $2048 \times 2048$ grid using various $\alpha's$. A maximum of ten subharmonic constellations were used for $\alpha$'s below  $\sfrac{11}{3}$, and a twenty subharmonic constellation maximum was used for $\alpha$'s greater than or equal to $\sfrac{11}{3}$.  $\mu_x=1$ and $\mu_y=2$ for all data points.}
\label{fig:Aniso Subh 2048}
\end{figure}

We have also included results for a $2048 \times 2048$ grid case, as shown in Fig. \ref{fig:Aniso Subh 2048} and in Table \ref{tab:min_error_2048}.  These metrics were gathered using a smaller number of total phase screens (5,000) due to the longer computation times associated with the larger grid sizes. For $\alpha \geq \sfrac{11}{3}$ we have included results up to $N_p=20$ in order to demonstrate that for higher $\alpha$'s performance improvements appear to continue as subharmonic constellations above 10 are added.  Results for select $\alpha$'s are shown in Figure \ref{fig:Aniso Subh 2048}, with 
We should note that due to the smaller number of total phase screens assessed, the statistical trends appear noisier than the $1024 \times 1024$ grid case.  Additionally, due to the overall lower error numbers for $\alpha < 3.8$ collected using the $1024 \times 1024$ grid over a greater number of sampled, we believe it is good assumption the higher resolution $2048 \times 2048$ grid results would be improved with a larger sampling phase screens under test.  For the $2048 \times 2048$ grid case, the average ratio of the randomized hybrid method's $\mathcal{E}$ to that of the traditional subharmonic method is 15.1\% across all $\alpha$'s under test.

\begin{table}
\begin{tabular}{ |c||p{3mm}|p{3mm}|p{3mm}|p{3mm}|p{3mm}|p{3mm}|p{3mm}|p{3mm}|p{3mm}|p{3mm}|p{3mm}|}
\hline
$\boldsymbol{\alpha}$  & \textbf{3.1} & \textbf{3.2} & \textbf{3.3} & \textbf{3.4} & \textbf{3.5} & \textbf{3.6} & $\boldsymbol{\frac{11}{3}}$ & \textbf{3.7} & \textbf{3.8} & \textbf{3.9} \\
\hline
$\boldsymbol{N_p}$ & 2 & 3  & 4 & 3 & 7 & 10 & 9 & 5 & 7 & 10 \\ 
$\boldsymbol{\mathcal{E}}$ & .65 & .33 & .82 & .86 & .26 & .46 & .61 & .55 & 1.9 & 16 \\ 
\hline
$\boldsymbol{N_p}$ & 1 & 2 & 2 & 2 & 2 & 3 & 3 & 4 & 5 & 9 \\
$\boldsymbol{\mathcal{E}}$  & 8.5 & 9.7 & 9.6 & 9.5 & 10 & 10 & 10 & 10 & 9.7 & 9.6 \\
\hline
\end{tabular}
\caption{Minimum percent RMS error ($\mathcal{E}$) and associated number of subharmonic constellations ($N_p$) for Randomized Hybrid Method (middle two rows) versus traditional method (bottom two rows) for $N_p \leq 10$. Data was collected over 50,000 phase screens for $1024 \times 1024$ grid, $\mu_x=1$ and $\mu_y=2$.}
\label{tab:min_error_1024}
\end{table}

\begin{table}
\begin{tabular}{ |c||p{3mm}|p{3mm}|p{3mm}|p{3mm}|p{3mm}|p{3mm}|p{3mm}|p{3mm}|p{3mm}|p{3mm}|p{3mm}|}
\hline
$\boldsymbol{\alpha}$  & \textbf{3.1} & \textbf{3.2} & \textbf{3.3} & \textbf{3.4} & \textbf{3.5} & \textbf{3.6} & $\boldsymbol{\frac{11}{3}}$ & \textbf{3.7} & \textbf{3.8} & \textbf{3.9} \\
\hline
$\boldsymbol{N_p}$ & 4 & 3 & 4 & 4 & 5 & 7 & 12 & 8 & 17 & 17 \\
$\boldsymbol{\mathcal{E}}$ & 1.4 & 1.3 & 1.1 & 1.5 & 1.3 & .95 & 1.5 & 1.2 & 1.3 & 4.0 \\
\hline
$\boldsymbol{N_p}$ & 1 & 2 & 2 & 2 & 2 & 3 & 3 & 4 & 5 & 10 \\
$\boldsymbol{\mathcal{E}}$  & 11 & 11 & 11 & 8.2 & 9.5 & 13 & 12 & 11 & 13 & 9.1 \\
\hline
\end{tabular}
\caption{Minimum percent RMS error ($\mathcal{E}$) and associated number of subharmonic constellations ($N_p$) for Randomized Hybrid Method (middle two rows) versus traditional method (bottom two rows). Data was collected over 5,000 phase screens for $2048 \times 2048$ grid, $\mu_x=1$ and $\mu_y=2$.}
\label{tab:min_error_2048}
\end{table}

\subsection{Effect of Direct Fourier Series on Computation Time}
\label{subsec:Effect of Direct Fourier Series on Computation Time}
In the previous two subsections we have demonstrated that the addition of subharmonics to the FFT-based screens is necessary to achieve accurate RMS error statistics for bounded spectral models with large outer scales, or unbounded spectral models with high power laws.  Although we have parameterized the simulation results so far using the number of subharmonic constellations in use, $N_p$, we note that the subharmonic modification can also be thought of as a non-uniform, discrete Fourier series with $N_f = 8 \cdot N_p$ frequency elements added to each FFT-based screen.  We have chosen to introduce $N_f$ in order to facilitate a comparison with methods using direct non-uniform Fourier series later in this subsection.  

When using subharmonic algorithms, it is quite noticeable to users that computation times are increased relative to the use of a purely FFT-based phase screen algorithm.  In order to quantify the computation time effects, we have designed a simulation routine where \textit{in-the-loop} computational delays are measured in a cumulative fashion over all trials.  By \textit{in-the-loop} delays, we are referring to computational steps which have to, by necessity, be performed each trial of the simulation, such as the creation of the random phase screens from trial to trial.  \textit{Out-of-the-loop} delays, on the other hand, do not need to be performed from trial to trial and in most cases do not have a considerable effect on computation time, and as such are not measured. For phase screen simulations which begin with a propagation of an undistorted Gaussian beam, as an example, we typically perform the propagation to the first screen just once outside of the Monte Carlo loop, meaning it is an \textit{out-of-the-loop} delay.  

\begin{figure}[ht!]
\centering\includegraphics[width=9cm]{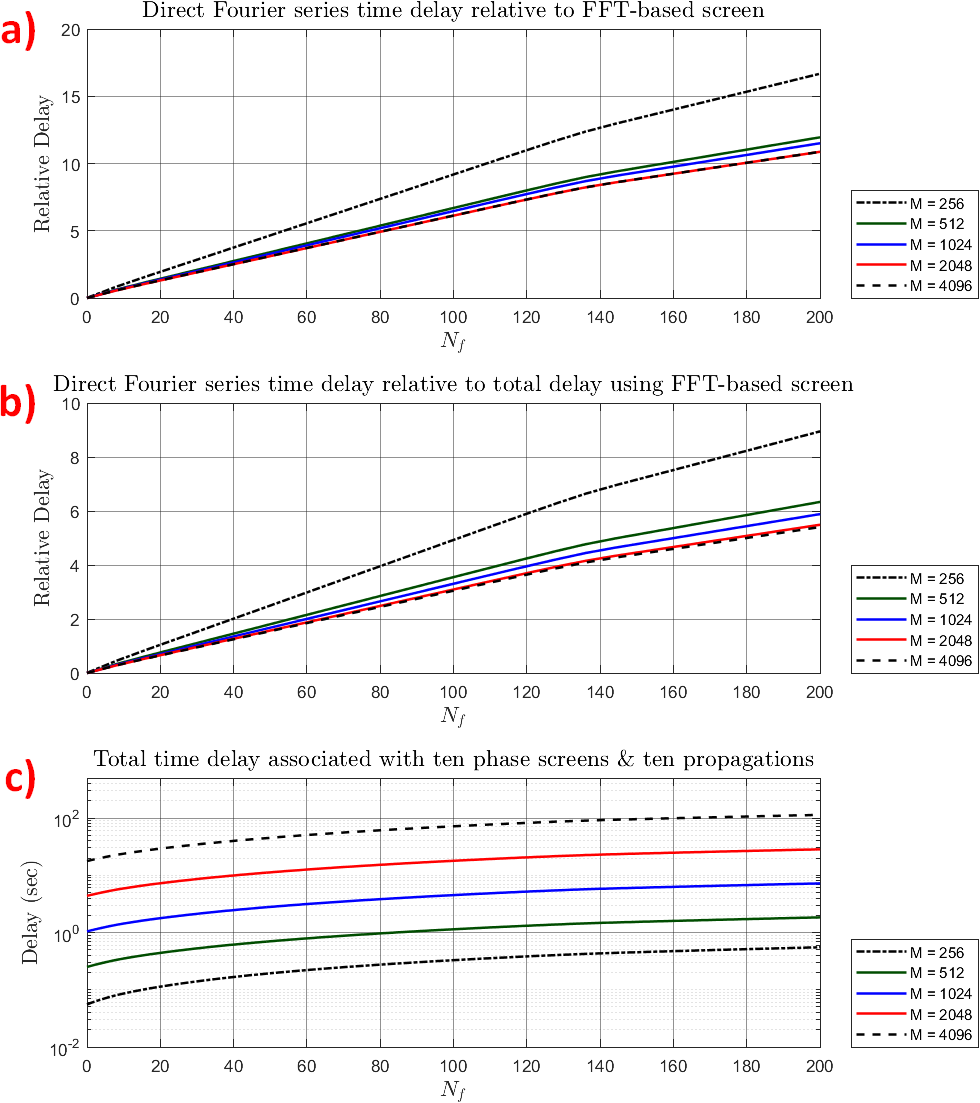}
\caption{a) Comparison of computation time delays associated with additional of a $N_f$ element non-uniform Fourier series relative to the delay associated with creating a purely FFT-base screen using the randomized method; b) Comparison of computation time delays associated with additional of a $N_f$ element non-uniform Fourier series relative to the total $in-the-loop$ computation time delay of a purely FFT-based simulation, including propagation; c) Absolute computation time delays associated with creation and application of ten phase screens, as well as ten propagation steps, for various $N_f$ values ($N_f=0$ denotes a purely FFT-based simulation).}
\label{fig:Computation Time}
\end{figure}

Although wave optics simulation parameters will vary depending on application, we have chosen nominal settings of ten screens and ten propagation stages for these measurements as these settings are somewhat typical for our purposes.  Simulations were performed in MATLAB using a Windows 10 computer equipped with a 3.6 GHz, 64-bit processor and 32 GB of RAM.  These tests utilized the unbounded Kolmogorov spectral model using twenty five subharmonic constellations ($N_p$).  One hundred simulation trials were run using grid sizes of 256 $\times$ 256, 512 $\times$ 512, 1024 $\times$ 1024, 2048 $\times$ 2048, and 4096 $\times$ 4096, and the various steps of the wave optics simulations were cumulatively timed for later comparison.  Each trial consisted of the creation of ten phase screens, phase modulation of the propagating wave using each phase screen, and ten propagations using the FFT-based angular spectrum propagation method \cite{Schmidt2010}.  For these computations, one real-valued phase screen derived from each complex screen was used in the simulation.  As the angular spectrum propagation method is designed for use with expanding and contracting spatial domains, we find that this is a common practice because two real-valued screens derived from one complex screen cannot be applied over differing spatial domains without additional interpolation steps.

Fig. \ref{fig:Computation Time} documents the results of the the simulations.  Subplot \ref{fig:Computation Time}a plots the growth in computation time necessary to produce each phase screen as a function of $N_f$ normalized by the time to produce a purely FFT-based screen, and averaged over one thousand screens.  Comparing the first and last points of the curves indicates rates of computation time increase ranging from 8.84\% per direct Fourier series component for the 256 $\times$ 256 grid to 5.94\% for the 4096 $\times$ 4096 grid.  Subplot \ref{fig:Computation Time}b places the results in the context of the overall computation time associated with a full wave optics simulation.  This subplot plots the growth in computation time associated with additional direct Fourier series components, as in subplot \ref{fig:Computation Time}a, but is normalized by the total \textit{in-the-loop} computation time of the wave optics simulation, which includes phase modulation of the simulated field with the phase screen distortion and propagation using the angular spectrum propagation algorithm.  The increases in total computation time ranged from 4.47\% per direct Fourier series component for the 256 $\times$ 256 grid to 2.70\% for the 4096 $\times$ 4096 grid.  Finally, Subplot \ref{fig:Computation Time}c shows the absolute times associated with each ten screen and ten propagation trial of the wave optics simulation, averaged over 100 trials.  We note that because the addition of white noise to the screens discussed in Subsection \ref{sec:Hybrid method for use with bounded and unbounded spectral models}\ref{subsec:Addition of White Noise to Phase Screens to Support Subresolution Inner Scales} may not be preferable for all spectral models we have not included this factor in the results shown, however we add that including this step increases the time associated with creation of the FFT-based screens by approximately 20\% for all grid sizes.  

As previously mentioned, the metrics presented in this subsection have been parameterized by $N_f$, the number of direct Fourier series components added to the screen using the subharmonic method, for the purpose of comparison with other methods.  Charnotskii \cite{Charnotskii2013} has previously designed methods for creation of atmospheric turbulence phase screens using randomized spectral sampling, which have been shown to produce phase screens with accurate statistics provided a sufficient number of non-uniform Fourier series elements are used \cite{Charnotskii2013_Statistics}. We note, however, that analysis for this method was performed using different spectral models than those used in this work, and also that RMS averaged or integrated statistics were not provided as functions of power law or inner-to-outer scale ratio. Additionally, the results of simulations in \cite{Charnotskii2013} and \cite{Charnotskii2013_Statistics} are compared to asymptotic approximations of the structure function at large and small scales, as opposed to the numerically computed (for bounded spectral models) or closed form solutions (for unbounded spectral models) as we have performed herein. These factors make a direct comparison between these methods rather difficult. We note in general, however, that in order to accurately approximate the structure function metrics of interest such that there is near overlap when plotted on the logarithmic scale a $N_f$ of 1,000 to 5,000 must be used.  

For the grid sizes which we have analyzed in this subsection, this number of discrete Fourier elements implies a significant increase in computation time relative relative to our method.  According to our analysis, the computation time difference associated with $N_f=1,000$ ranges from 27.0 times larger than that of the purely FFT-based simulation for the 4096 $\times$ 4096 grid, to 44.7 to times larger than that of the purely FFT-based simulation for the 256 $\times$ 256 grid.  Although our randomized algorithm without the use of subharmonics may produce acceptable results for many purposes, in Subsections \ref{sec:Hybrid method for use with bounded and unbounded spectral models}\ref{subsec:Randomized_FFT-based_Sampling_Results} and \ref{sec:Hybrid method for use with bounded and unbounded spectral models}\ref{subsec:Results for Unbounded Spectral Models} we have demonstrated that our hybrid algorithm produces very accurate results for $N_p \leq 10$, $N_f \leq 80$ for all but one case, which was for non-Kolmogorov turbulence with $\alpha=3.9$.  Comparing the difference between the total delay associated with a wave optics simulation using the $N_f = 80$ hybrid screens (including FFT-based component) and the extrapolated delay associated with a non-FFT $N_f = 1,000$ screen predicts that the latter method would result in delays 7.84 times higher for the 4096 $\times$ 4096 grid, and 9.04 times higher for the 256 $\times$ 256 grid.  Obviously, we assume that the delays associated with $N_f=5,000$ would be five times higher than for the $N_f=1,000$ cases, which may lead to unacceptable computation time delays for larger grid sizes.

\section{Conclusion and Discussion}
\label{sec:Conclusion}
In this work we have demonstrated a straightforward method to correct low spatial frequency and periodicity issues associated with FFT-based phase screen methods.  We have further demonstrated that low and high spatial frequency contributions can be further improved by combining our core technique with supplemental methods.  In previous work \cite{Paulson2018} we have estimated that the addition of every subharmonic constellations adds approximately the same computation time of generating the purely FFT-based phase screen.  As in many cases the core algorithm sans subharmonics outperforms other methods utilizing multiple subharmonics, significant computation time associated with creating the screens may be conserved while garnering improved structure function accuracy.  In cases where accuracy is paramount, our hybrid method is able to produce very low error percentages across the region of interest.

Aside from the stated application of simulation of atmospheric optical turbulence, utilizing randomized spectral sampling in concert with the FFT may have additional applications.  When modelling processes containing divergences or nulls in their spectral representations we believe a modification of our algorithm giving in Section \ref{sec:Randomized_FFT-based_Sampling}\ref{subsec:Randomized_FFT-based_Sampling_Algorithm} should be considered.  Simulations of processes approximated by fractional differencing \cite{HOSKING1981}, such as radio frequency oscillator phase noise \cite{Leeson2016}, are a potential candidate. The $1/f^{\ \alpha}$ spectral model (with $f$ the time frequency) of the stochastic process is similar to our own.  Additionally, straightforward applications would be to apply the techniques outlined in this study to simulations of partially coherent sources \cite{Schell1967,Dogariu:03}, simulations using three-dimensionally correlated phase screens \cite{Naeh:14,FATHEDDIN2017}, and simulations of optical propagation in underwater turbulence, which has its own unique spectral representations \cite{Goodman1990,Korotkova2012,Hou:13}.  The heightened low spatial frequency accuracy our techniques will have a direct impact on beam wander statistics, which have been shown as an important factor in free space optical (FSO) communications \cite{Dios:04,Andrews2006,Basu:08}.  Additionally, temporal statistics are often at the center of studies regarding FSO performance metrics \cite{Yura:83,Zhu2002,Zhu2003,Basu:08}, which has lead to research on long phase screens to simulate aperiodic turbulence moving at the wind speed \cite{Dios:04,Assemat:06,Vorontsov2008}.  We note that through a combination of circularly shifting the $C$ matrix of Eq. \ref{eq:C_def} and applying multiplying the exponential term in Eq. \ref{eq:dither_phase_screen_FFT} by a factor, we have successfully demonstrated the creation of aperiodic moving phase screens derived from the original components of $\theta_R$, but not requiring the use of additional FFT's after the first complex screen realization.  Extending this capability to include subharmonic constellations is not challenging.  This feature can be directly applied to simulations of time-domain turbulence affects, or combined with other techniques \cite{Vorontsov2008,Johnston:00} to improve statistics.

\section{Funding Information}
This work was supported by the Office of Naval Research (ONR) Atmospheric Propagation Studies for High Energy Lasers (APSHEL) program under grant N000141812008.

\bibliographystyle{unsrt}  
\bibliography{MAIN}  %%% Remove comment to use the external .bib file (using bibtex).


% George Kour and Raid Saabne.
% \newblock Real-time segmentation of on-line handwritten arabic script.
% \newblock In {\em Frontiers in Handwriting Recognition (ICFHR), 2014 14th
%   International Conference on}, pages 417--422. IEEE, 2014.

% George Kour and Raid Saabne.
% \newblock Fast classification of handwritten on-line arabic characters.
% \newblock In {\em Soft Computing and Pattern Recognition (SoCPaR), 2014 6th
%   International Conference of}, pages 312--318. IEEE, 2014.

% \bibitem{hadash2018estimate}
% Guy Hadash, Einat Kermany, Boaz Carmeli, Ofer Lavi, George Kour, and Alon
% \newblock Estimate and replace: A novel approach to integrating deep neural
%   networks with existing applications.
% \newblock {\em arXiv preprint arXiv:1804.09028}, 2018.

% \end{thebibliography}

\end{document}